\documentclass[10pt]{article}

\usepackage[usenames,dvipsnames,svgnames,table]{xcolor} % usenames : 16개 기본색,  dvipsnames : 또 다른 68개 색, svgnames : 또 다른 150개 색에 접근할 수 있다. \usepackage{color} 대신 쓰인다. xcolor 패키지 설명서에서 색깔을 찾아볼 수 있다.
\usepackage{kotex}
\usepackage{graphicx}
\usepackage{epsfig}
\usepackage{amssymb, amsmath, mathtools}
\usepackage{verbatim}
\usepackage{harvard} %dcu를 쓰려면 이것을 써야한다.
\usepackage[affil-it]{authblk}
\usepackage{mathrsfs}
\usepackage{here}
\usepackage{makecell}
\usepackage{tikz}
\usepackage{enumerate}
\usepackage[shortlabels]{enumitem}
\usepackage{yfonts}
\usepackage{comment}
\usepackage{longtable}

\usepackage{hyperref}
\usepackage[colorinlistoftodos, textsize=scriptsize]{todonotes} % 노트를 만들기 위한 패키지. disable 옵션을 쓰면 노트를 모두 없애준다.

\hypersetup{
  colorlinks,
  citecolor=Black,
  linkcolor=Black,
  urlcolor=Black} % 인용색을 검은색으로 정의.

\usepackage[framemethod=TikZ]{mdframed}
\mdfdefinestyle{JL}{%
    linecolor=blue,
    outerlinewidth=1pt,
    roundcorner=2pt,
    innertopmargin=0.1\baselineskip,
    innerbottommargin=0.1\baselineskip,
    innerrightmargin=2pt,
    innerleftmargin=2pt,
    backgroundcolor=orange!100!white}
\newenvironment{JL}{\begin{mdframed}[style=JL]  \footnotesize} { \end{mdframed}}
\newcommand{\bJ}{\begin{JL}}
\newcommand{\eJ}{\end{JL}}

\includecomment{note} %\begin{note} \end{note} 혹은 \note \endnote로 된 부분을 포함한다.
\newtheorem{theorem}{Theorem}[section]
\newtheorem{lemma}[theorem]{Lemma}
\newtheorem{definition}[theorem]{Definition}
\newtheorem{proposition}[theorem]{Proposition}
\newtheorem{corollary}[theorem]{Corollary}

\newtheorem{remark}[theorem]{Remark}

\everymath{\displaystyle}

\newcommand{\half}{\frac{1}{2}}

\newcommand{\iid}{\stackrel{i.i.d.}{\sim}}

\newcommand{\be}{\begin{equation}}
\newcommand{\ee}{\end{equation}}
\newcommand{\bea}{\begin{eqnarray*}}
\newcommand{\eea}{\end{eqnarray*}}
\newcommand{\bean}{\begin{eqnarray}}
\newcommand{\eean}{\end{eqnarray}}
\newcommand{\ben}{\begin{enumerate}}
\newcommand{\een}{\end{enumerate}}
\newcommand{\bi}{\begin{itemize}}
\newcommand{\ei}{\end{itemize}}
\newcommand{\brem}{\begin{remark}}
\newcommand{\erem}{\end{remark}}
\newcommand{\bcen}{\begin{center}}
\newcommand{\ecen}{\end{center}}
\newcommand{\bsv}{\begin{semiverbatim}}
\newcommand{\esv}{\end{semiverbatim}}

\newcommand{\bt}{\begin{theorem}}
\newcommand{\et}{\end{theorem}}
\newcommand{\bl}{\begin{lemma}}
\newcommand{\el}{\end{lemma}}
\newcommand{\bd}{\begin{definition}}
\newcommand{\ed}{\end{definition}}
\newcommand{\bc}{\begin{corollary}}
\newcommand{\ec}{\end{corollary}}
\newcommand{\bp}{\begin{proposition}}
\newcommand{\ep}{\end{proposition}}

\newcommand{\bff}{ \mathbf{f}}

\newcommand{\bfx}{ \mathbf{x}}
\newcommand{\bfy}{ \mathbf{y}}

\newcommand{\bsigma}{ \boldsymbol{\sigma}}

\title{Ordinary Differential Equation Models and their Computation Methods}
\author{Jaeyong Lee}
\affil{Department of Statistics \\ Seoul National University}

\begin{document}

\maketitle
\setcounter{tocdepth}{2}
%\tableofcontents

\begin{abstract}
In this article, I introduce the differential equation model and review their frequentist and Bayesian computation methods. A numerical example of the FitzHugh-Nagumo model is given. 
\end{abstract}

\section{Introduction}
I spent a Sabbatical year in the Statistical and Applied Mathematical Sciences Institute in 2010. During the Sabbatical year, I attended a talk given by James Ramsay. He gave a talk on the differential equation model and an example of modeling underground water levels in some place in Canada.  The water level data  can be seen as a function of time and its plot was not smooth at all. It was wiggly and has big jumps. The water level was mostly affected by the rainfall, but with some time lags. 
He showed the fitted curves  and predictions. I was very surprised  that his fitted curve as well as the predictions follow the observed data very closely. I would not be surprised by the fitted curve following the data closely if the model has thousands of parameters, but his estimate follows the data with just a handful of parameters. At the talk I imagined myself analyzing the same data set without the differential equation model. I thought about many complicated nonparametric models, but I could not think of a statistical model which would resemble the fitted curve he produce on his slide. This is how I got drawn to the topic of the differential equation model.  \\

The differential equation is a primary mathematical model that describes dynamical systems which are first developed by Henri Poincar\'e to explain celestial bodies. Differential equations are widely used in many different areas such as the epidemic model, climate model, economic model, and chemical engineering, to name just a few.  The statistics community however largely neglected the differential equation modeling until recently. \\

The differential equation model is a statistical model that consists of two equations: 
\begin{align}
\text{observation equation: } & & y(t_i) & = x(t_i) + \epsilon(t_i), ~i=1,2, \ldots, n,  \label{ob.eq}\\
\text{differential equation: } & & \frac{d x(t) }{dt} & = f(x(t), t; \theta),  \label{diff.eq}
\end{align}
where $0 < t_1 < t_2 < \ldots < t_n < T$ are time points where observations $y(t)$ are observed and $\epsilon(t_i)$'s are observational error typically assumed to follow $N(0, \Sigma)$ with positive definite matrix $\Sigma$. The error variance $\Sigma$ is assume typically a diagonal matrix $\Sigma = diag(\sigma_1^2, \ldots, \sigma_p^2)$. Here $y(t)$ is a $p$-dimensional vector that are observable and $x(t)$ is also a $p$-dimensional vector function that satisfies the differential equation \eqref{diff.eq} and it is the mean function of the regression model \eqref{ob.eq}. The function $f$ is smooth and uniquely determines $x$ when the initial value of $x$, $x_0 = x(0)$ is given. Note that the differential equation model \eqref{ob.eq}  and \eqref{diff.eq} is just a nonlinear regression model where the mean function of the regression model is expressed as a solution of a differential equation. It also resembles the state-space model. One can view that the differential equation model as the state-space model whose state equation is expressed as a differential equation. \\

Here I give an example of the differential equation model. The SIR model is the differential equation model for disease spread in a small community. It consists of differential equations for three curves, susceptible (S), infected (I), and recovered (R): 
\begin{eqnarray*}
\frac{dS(t)} {dt} & = & -\beta \frac{1}{N} I(t) S(t) \\
\frac{dI(t)} {dt} & = & \beta \frac{1}{N} I(t) S(t) -\gamma I(t) \\
\frac{dR(t)} {dt} & = & \gamma I(t), 
\end{eqnarray*}
where $t$ is time in $[0, T]$, for some $T> 0$, $N$ is the total population of the community, and $S(t)$,  $I(t)$ and $R(t)$ are 
the numbers of the susceptibles, the infected and the recovered at time $t$, respectively. There are three equations in the model, but fundamental quantities are two terms. One is $\beta \frac{1}{N} I(t) S(t) $ which   can be interpreted as the number of infected people in a unit time. The parameter $\beta' = \beta/N$ is the probability that  one infected and one susceptible meet and the susceptible gets infected during a unit time and $\beta$ is the number of people in the population or among $N$ susceptibles who get infected by one infected during a unit time. Some authors prefer  $\beta'$ and use 
$$\frac{dS(t)} {dt}  =  -\beta' I(t) S(t).$$ 
Another fundamental quantity is $\gamma I(t)$ which is the number of recovered during a unit time. The parameter $\gamma$ is the recovery rate of the infected during a unit time. For the SIR model, we assume the population size $N$ remains constant over the time period $[0, T]$ and 
$$S(t) + I(t) + R(t) = N, ~ \forall t.$$
Thus, essentially we need only two equations. \\

In the differential equation model, we assume that we do not observe $S(t)$, $I(t)$ and $R(t)$ directly, but observe them with errors, i.e. we observe $S^*(t_i)$ and $I^*(t_i)$ at $0 < t_1 < t_2 < \ldots < t_n < T$: 
\begin{eqnarray*}
S^*(t_i) & = & S(t_i) + \epsilon_1(t_i), \\
I^*(t_i) & = & I(t_i) + \epsilon_2(t_i),
\end{eqnarray*}
where $\epsilon_1(t)$ and $\epsilon_2(t)$ are errors. \\

This article is organized as follows. In sections 2 and 3, frequentist and Bayesian methods in the literature are reviewed, respectively. In section 3, a numerical example of the FitzHugh-Nagumo model is given. 

\section{Frequentist Methods} 
\subsection{Explicit Numerical Integration of Differential Equation}
\citeasnoun{bard1974nonlinear} considers in his book the parameter estimation of the dynamic model with errors. 
The parameter of the differential equation is estimated with the least squares method, i.e., 
$$\hat{\theta} := argmin_{\theta} ||y_i - x(t_i) ||^2.$$
In the course of the optimization, whenever a value of  $x(t)$ is required, a numerical solver is invoked. 

This method performs well when the sample size and the number of parameters are small. But its computation can be prohibitive when the model gets larger. When the numerical solver is the 4th order Runge-Kutta algorithm, the estimator is asymptotically efficient (Xue et al. 2010, Bhaumik and Ghosal 2017). \nocite{xue2010sieve}  \nocite{bhaumik2017efficient}

\subsection{Two-step methods} 
\citeasnoun{varah1982spline} proposed the two-step method. In the first step, $x(t)$ is estimated with a cubic spline with fixed knots using the observed data  $\bfy = (y(t_1), \ldots, y(t_n) )$ without considering the differential equation. In the second step, $\theta$ is obtained by minimizing the distance between the estimated $\frac{d\hat{x} (t)}{dt}  $ and $f(\hat{x}(t), t, \theta)$, i.e., 
\begin{equation} \label{two_step}
\hat{\theta} := argmin_\theta \sum_{i=1}^n ||\frac{d\hat{x} (t_i)}{dt} - f(\hat{x}(t_i), t_i, \theta) ||^2.
\end{equation}

There are variations in the first and the second steps with different nonparametric function estimation methods and different norms. This method is not asymptotically efficient but computationally fast. 

\subsection{Iterated Principal Differential Analysis} 
\citeasnoun{ramsay2005functional} considered a variation of two-step method. In the proposed method, the following two steps are iterated. 

In the first step, the cubic spline with fixed knots is fitted but with a penalty  term, i.e., 
$$\hat{x} = argmin_x \sum_{i=1}^n ||y_i - x(t_i) ||^2 + \lambda \int (\dot{x} (t) - f(x, t, \theta) )^2 dt,$$
where $\theta$ is the estimated value in the previous iteration. 
In the second step, $\theta$ is obtained by minimizing \eqref{two_step}. 

\subsection{Generalized Profiling Method} 
\citeasnoun{ramsay2007parameter} proposed the generalized profiling method. They grouped the parameters in three groups, regularization parameter $\lambda$, the parameter in the differential equation $\theta$ and $\sigma$'s, and  regression coefficients $\beta$ of basis expansion of x. Each group is estimated and thus eliminated in turn in three steps.

In the first step, $\lambda$ is estimated by a generalized cross-validation criterion. In the first step, whenever one needs values of $\theta$, $\sigma$'s and $\beta$, one evaluate them in the following second and third step.  In the second step, $\theta(\lambda)$ is estimated by minimizing 
$$\hat{\theta}(\lambda) := argmin_{\theta} ||y_i - x(t_i) ||^2$$ 
where $x(t) = \beta' b(t)$, $b(t) = (b_1(t), \ldots, b_k(t))$ and $b_j(t)$ are the  B-spline basis. 
In the third step, $\beta(\lambda, \theta)$ is estimated by minimizing 
$$ \sum_{i=1}^n ||y_i - x(t_i) ||^2 + \lambda \int (\dot{x} (t) - f(x, t, \theta) )^2 dt,$$
where  $\dot{x}(t) = \beta' \dot{b}(t)$. In the above description, $\theta(\lambda)$ and $\beta(\lambda, \theta)$ are used to emphasize the dependence to $\lambda$, and $\lambda$ and $\theta$, respectively, The generalized profiling estimator is asymptotically efficient \cite{qi2010asymptotic}.

\section{Bayesian Methods}
In this section, the Bayesian estimation methods for the differential equation model are reviewed. 
There are three groups of parameters in the differential equation model. Two are obvious and one is not. The two groups of  obvious parameters are the parameter in the differential equation \eqref{diff.eq}, $\theta$, and the observational error variance $\Sigma$.  Under some smoothness conditions, the differential equation together with initial value of $x$, $x_0 := x(0)$, uniquely determine $x$. Thus, one needs to include  the initial value of $x$, $x_0 := x(0)$, which is not so obvious at a glance. To complete the Bayesian model, we need to put  a prior on $\theta, \Sigma$ and $x_0$. We will denote the prior 
$$(\theta, \bsigma^2, x_0) \sim \pi(\theta) \pi(\bsigma) \pi(x_0).$$

\subsection{Bayesian Method with Explicit Integration of Differential Equation}
The first approach was the approach taken in \citeasnoun{gelman1996physiological}. The posterior is 
$$\pi(\theta, \bsigma^2, x_0 | \bfy, \bfx) \propto \pi(\theta, \bsigma^2, x_0) \times p(\bfy | \bfx, \bsigma, x_0),$$
where
$$p(\bfy | \bfx, \bsigma, x_0)  = \prod_{i=1}^n 
| 2\pi \Sigma |^{-\half} e^{ - \frac{1}{2}  (y_i - x(t_i; \theta, x_0) )' \Sigma^{-1} (y_i - x(t_i; \theta, x_0) )},$$
 $x(t)$ is denoted as $x(t; \theta, x_0)$ to show the dependence on $\theta$ and $x_0$, $\bfx = (x(t_1), \ldots, x(t_n))$, $\bfy = (y(t_1), y(t_2), \ldots, y(t_n))$ and $\bsigma = (\sigma_1, \sigma_2, \ldots, \sigma_p)$. To compute the posterior, one can apply the Markov chain Monte Carlo  algorithm such as Metropolis-Hastings sampler or Hamiltonian Monte Carlo. The advantage of this method  is that it computes the exact posterior while other methods employ certain approximations to ease the posterior computation. 
The difficulties with this approach are twofold. First,  the analytic solution of the differential equation \eqref{diff.eq} is typically not available and whenever you need the likelihood, you need to resort to a numerical solver of the differential equation. This makes the posterior computation intensive. Second, the differential equation can be chaotic and a small variation in the parameter may result in completely different curves. Thus, estimation of the parameter can be very sensitive and it affects at times poor performances of the estimator. 

\citeasnoun{dass2017laplace} used the Laplace approximation method to speed up the posterior computation in which the function $x$ is evaluated using mathematical formulas of the Euler method or the 4th order Runge-Kutta method.  This method gives accurate parameter estimates with fast computing time with a moderate dimension sizes of $\theta$. But when the dimension of $\theta$ is high, the computation can be intensive. 

\citeasnoun{bhaumik2017efficient} proved the Bernstein-von Mises theorem and the posterior is asymptotically efficient when the numerical solver is the 4th order Runge-Kutta.

\subsection{Bayesian Collocation Method}
%There are attempts solve this problem. 
\citeasnoun{campbell2012smooth} use a Bayesian approach based on the collocation method and a prior on $x(t)$  was based on the penalty. The collocation approach is to express $x$ with a linear combination of bases, i.e., 
$$x(t) = \sum_{j=1}^k \beta_j b_j(t),$$
where $b_j(t)$'s are bases of function spaces, e.g. spline bases. 
The model can be viewed  as 
$$y(t_i)  =  x(t_i) + \epsilon(t_i), ~ \epsilon(t_i) \iid N(0, \sigma^2), ~ i=1,2, \ldots, n$$
and  $\frac{dx}{dt} = \dot{x}$ has prior 
$$exp[ - \lambda PEN(x)],$$
where
$$PEN(x) = \int || \frac{dx}{dt} - f(x(t), t, u(t); \theta) ||^2 dt.$$
Thus, the posterior is 
$$\pi(\theta, \sigma^2, \alpha) \propto e^{-\lambda PEN(x; \alpha, \theta)} \times p(\bfy | \bfx, \bsigma, x_0) . $$
The approach taken by \citeasnoun{campbell2012smooth} is slightly more complicated than the above and used tempering idea with different $\lambda$s.

\subsection{Gaussain Process Approaches}
The key fact used in the Gaussian process approach is the fact  that when $x$ follows a  Gaussian process, $x$ and $\dot{x}$ are jointly a Gaussian process. This contradicts the mathematical fact that when $x$ is a function, $x$ determines $\dot{x}$. Nevertheless, it gives a methodological advantage in ordinary differential equation (ODE) models. 
There are two approaches in this category: adaptive gradient matching (AGM) method (Calderhead et al. 2009; Dondelinger et al. 2013)\nocite{calderhead2008accelerating}\nocite{dondelinger2013ODE}  and Gaussian process ODE (GPODE) method \cite{wang2014gaussian}. \\

Suppose $x \sim GP(0, C_\phi)$, the Gaussian process with mean function $0$ and covariance function $C_\phi$  with parameter $\phi$. 
In the AGM method, the posterior is obtained by 
$$\pi(\theta, \bsigma, \phi | \bfx, \bfy) \propto p(\bfy | \bfx, \bsigma, x_0) \times \pi(x|\theta, \phi, \gamma) \times \pi(\theta, \phi, \gamma, \bsigma),$$
where $\pi(\theta, \phi, \gamma, \bsigma)$ is the prior  $(\theta, \phi, \gamma, \bsigma)$ and $\gamma$ is the additional error variance introduced to $\dot{x}$ in \eqref{ssm1}. 
The prior  $\pi(x|\theta, \phi, \gamma)$ is constructed by 
\begin{equation}\label{gp1}
\pi(x | \theta, \phi, \gamma) = \int \pi(x, \dot{x} | \theta, \phi, \gamma) d\dot{x}
\end{equation}
where 
\begin{equation}\label{gp2}
\pi(x, \dot{x} | \theta, \phi, \gamma) \propto GP(x, | 0, C_\phi) GP(\dot{x} | x, \phi) \prod_{i=1}^n N(\dot{x}(t_i) | f(t_i, x(t_i), \theta), \gamma I).
\end{equation}
Note that  integration \eqref{gp1} can be performed explicitly. In \eqref{gp2}, the second factor $GP(\dot{x} | x, \phi)$ of the right hand side is derived from the assumption that $x$ follows a Gaussian process and the third factor by relaxing  differential equation \eqref{diff.eq} with 
\begin{equation}\label{ssm1}
\dot{x}(t_i) = f(x(t_i), t_i; \theta) + \nu_i, ~ \nu_i \iid N(0, \gamma).
\end{equation}
Note that this is not a legitimate derivation following probability laws, because the distributions of $\dot{x}$ from two sources are multiplied.   \\

In the GPODE method, the posterior is obtained as 
$$\pi(\theta, \phi, \bsigma | \bfy, \bfx) \propto p(\bfy | \bff, \phi, \bsigma) \times GP(x | 0, C_\phi),$$
where $\bff = (f(t_i, x(t_i), \theta))_{i=1}^n$ and $p(\bfy | \bff, \phi, \bsigma)$  is $p(\bfy | \dot{\bfx}, \phi, \bsigma) $ with $ \dot{\bfx}$ replaced by $\bff$ using the differential equation. The conditional distribution $p(\bfy | \dot{\bfx}, \phi, \bsigma) $ is obtained by 
$$p(\bfy | \dot{\bfx}, \phi, \bsigma ) = \int p(\bfy | \dot{\bfx}, \phi, \bsigma)  \times GP(\bfx | \dot{\bfx}, \phi) d\bfx.$$
Conveniently, this integration also can be done explicitly under the normal error assumption. 

\subsection{Two-Step Approaches}
\citeasnoun{bhaumik2015bayesian} considered two step approach to the posterior computation of the ODE model. In the first step, the regression function $x$ is estimated without considering the differential equation using nonparametric Bayesian model with the B-spline basis
$$ y_j = x(t_j) + \epsilon(t_j), j=1,2, \ldots, n$$
where $x(t) = \sum_{j=1}^k \beta_j b_j(t)$, $b_j(t)$ are the B-spline basis and $\beta_j$ are coefficients of the basis. As a result of the first step, the posterior of $x(t)$ or $\beta_j$'s are obtained.

In the second step, one sample $\dot{x}$ from the posterior of $x$. Since the derivatives of the B-spline basis are well-known, one first samples $x$ from the posterior and can compute $\dot{x}$ analytically. For each sample of $\dot{x}$, one finds the matched $\theta$ by minimizing 
$$\theta^* := argmin_\theta \int || \dot{x}(t) - f(x(t), \theta) ||^2 w(t) dt$$
where $w(t)$ is a weight function. The posterior of $\theta$ is approximated by collecting $\theta^*$'s. \citeasnoun{bhaumik2015bayesian} proved the posterior satisfies the Bernstein-von Mises theorem. 

Two step approaches are not generally asymptotically efficient. But \citeasnoun{bhaumik2017efficient} considered the second step where the posterior sample $\theta^*$ is obtained by matching the Runge-Kutta numerical solution as follows: 
$$\theta^* := argmin_\theta \int || x(t) - x^*_\theta(t) ||^2 dt,$$
where $x(t)$ is the posterior sample from the first step and $x^*_\theta(t) $ is the solution of the Runge-Kutta method with parameter $\theta$. They proved that the Bayes estimator from this posterior is asymptotically efficient.

\subsection{Nonlinear State-Space Model Approximation} 
The relaxed differential equation model (RDEM) \cite{lee2018inference} approximated the differential equation model \eqref{diff.eq}, and the differential equation model is approximated by the  nonlinear state-space model 
\begin{eqnarray}
y_i & = & x_i + \epsilon_i, ~ \epsilon_i \iid N(0, \Sigma) \\ \label{ssm2}
x_{i} & = & g(x_{i-1}, t_{i-1}; \theta) + \eta_i, ~ \eta_i \iid N(0, V), ~ i = 1, \ldots, n, \label{ssm3}
\end{eqnarray}
where
\bea
g(x_i, t_i; \theta) & = & x_i + {h_{i+1} \over 6}(k_{i1} + 2k_{i2} + 2k_{i3} + k_{i4}) \\
k_{i1} &=& f(x_i, t_i ; \theta) , \\ 
k_{i2} &=& f(x_i + {h_{i+1} \over 2}k_{i1} , t_i + {h_{i+1}\over2} ;\theta), \\ 
k_{i3} &=& f(x_i + {h_{i+1} \over2}k_{i2},t_i + {h_{i+1} \over2} ;\theta), \\ 
k_{i4} &=& f(x_i + k_{i3} , t_i + h_{i+1};\theta)
\eea
and  $h_{i+1} = t_{i+1} - t_i$. 
Note that $g(x_i, t_i; \theta)$ in \eqref{ssm3} is the equation of the 4th order Runge-Kutta method. 

The posterior computation is done by a sequential Monte Carlo Method, the extended Liu and West filter \cite{rios2013extended} in which $x_i$ is further relaxed with added noises. In the numerical experiments,  RDEM saves the computation time drastically while the accuracy is comparable to the other methods. 

The SSVB (state-space model with variational Bayes) \cite{yang2021variational} applies the variational Bayes method to obtain the posterior of \eqref{ssm2} and \eqref{ssm3}. \citeasnoun{yang2021variational} considered the Lorenz-96 model \cite{lorenz1995predictability} with $p=10$ variables as a numerical testbed: 
 $$  \frac{dX_j}{dt} = (X_{j+1}-X_{j-2})X_{j-1}-X_j+F,\qquad\text{for }\ j=1,\  \dots,\ p,$$
and according to the circular structure, $X_{-1}=X_{p-1}$, $X_0=X_p$, and $X_{p+1}=X_1$.
The Lorenz-96 model is a nice numerical testbed for the differential equation model, for it can be expanded as desired by increasing $p$. 
For the Lorenz-96 model with $10$ variables, none of the competitors can estimate the true parameters reasonably  while the SSVB performed reasonably well. \\

The SSVB is computationally fast and gives numerically stable estimates, but inherits the lack of variance estimates from the variational method. \citeasnoun{yang2021laplace} proposed to estimate the variance of the posterior using the Laplace method.

\section{FitzHugh-Nagumo Model} 
The FitzHugh-Nagumo model (FitzHugh, 1961\nocite{fitzhugh1961impulses}; Nagumo et al. 1962\nocite{nagumo1965active}) describes the action of spike potential in the giant axon of squid neurons by an ODE with two state variables and three parameters:
\bea
\dot{x}_1(t) &=& \theta_3 \left( x_1(t) - \frac{1}{3}x_1^3(t) + x_2(t) \right) , \\
\dot{x}_2(t) &=& -\frac{1}{\theta_3} \Big( x_1(t) - \theta_1 + \theta_2 x_2(t)  \Big),
\eea
where $-0.8 < \theta_1, \theta_2 < 0.8$ and $0 < \theta_3 <8$. The two state variables, $x_1(t)$ and $x_2(t)$, are the voltage across an membrane and outward currents at time $t$, respectively. 

\begin{figure*}[!bt]
	\centering
	\includegraphics[width=16cm,height=10cm]{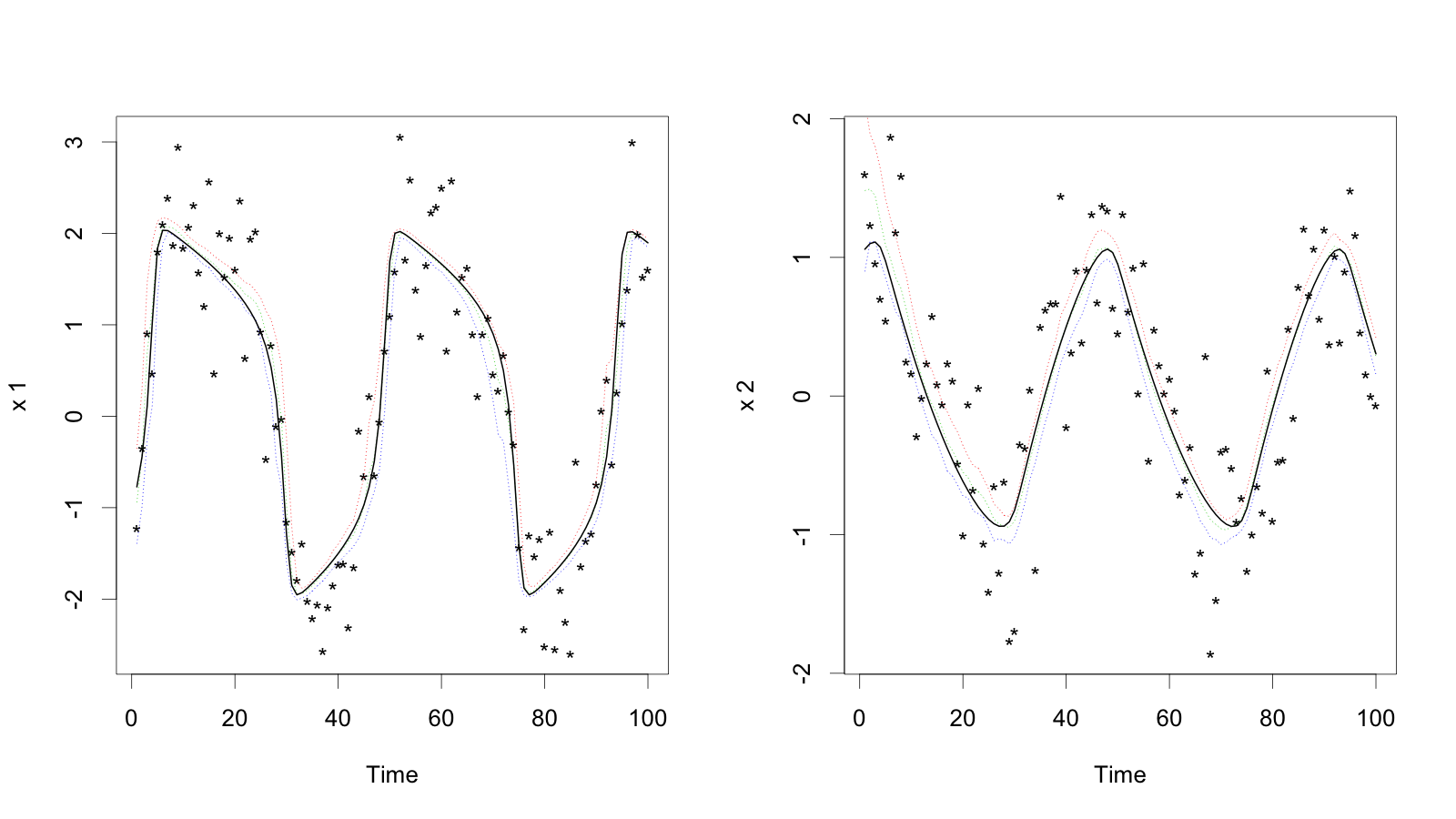}
	\caption{True and estimated curves of FitzHugh-Nagumo model. 
		}
	\label{fig:fwdbwd}
\end{figure*}

The data are generated from the differential equation model with the differential equation being the FitzHugh-Nagumo model and the fitted curves are given in Figure \ref{fig:fwdbwd}. In this particular example, the RDEM of \citeasnoun{dass2017laplace} is used. The solid lines are $x_1(t)$ and $x_2(t)$ as a function of time from the FitzHugh-Nagumo model with $x(t_0) = (-1, 1)^T, \theta = (0.2, 0.2, 3)^T$. The star-shaped points are the generated data of the populations with $\sigma^2=0.25$.
		The upper, lower and middle dotted lines are the 95 and 5 \% quantiles and median of the posterior $\pi(x_i \mid \bfy)$, respectively.

\bibliographystyle{dcu}
\bibliography{ode}

\end{document}